\documentclass[submission,copyright,creativecommons]{eptcs}

\usepackage{breakurl}             % Not needed if you use pdflatex only.
\usepackage{underscore}           % Only needed if you use pdflatex.
\usepackage[english]{babel}
\usepackage{times}
\usepackage{helvet}
\usepackage{courier}
\usepackage{amsmath}
\usepackage{amssymb}
\usepackage{enumerate}
\usepackage{graphicx}
\usepackage{url}
\usepackage{listings}

\def\titlerunning{Strong Equivalence for  $\lpmln$ Programs}
\def\authorrunning{Joohyung Lee and Man Luo}

\long\def\BOC#1\EOC{\message{(Commented text )}}
\long\def\BOCC#1\EOCC{\message{(Commented text )}}
\long\def\BOCCC#1\EOCCC{\message{(Commented text )}}
\long\def\optional#1{\empty}
\long\def\NBB#1{}

\def\o{\overline}
\def\ar{\leftarrow}
\def\bi{\begin{itemize}}
\def\ei{\end{itemize}}
\def\beq{\begin{equation}}
\def\eeq#1{\label{#1}\end{equation}}
\def\ba{\begin{array}}
\def\ea{\end{array}}
\def\i#1{\hbox{\it #1\/}}

\def\sm{\rm SM}

\def\no{\i{not}}

\def\ar{\leftarrow}
\def\rar{\rightarrow}
\def\lrar{\leftrightarrow}
\def\no{\i{not}}

\def\i#1{\hbox{\itshape #1\/}}

\def\lpmln{{\rm LP}^{\rm{MLN}}}
\def\modelsht{\mathrel{\mathop{\models}_{\!\!\!\!\!{\scriptscriptstyle\rm ht}}}}

\def\:{\!:\!}

\def\bi{\begin{itemize}}
\def\ei{\end{itemize}}

\def\mo#1{{\mathsf{#1}}}

\newtheorem{prop}{Proposition}
\newtheorem{thm}{Theorem}

\newtheorem{definition}{Definition}
\newtheorem{example}{Example}
\newtheorem{lemma}{Lemma}

\title{Strong Equivalence for  $\lpmln$ Programs}

\author{Joohyung Lee and Man Luo 
\institute{Arizona State University, USA}
\email{\{joolee, mluo26\}@asu.edu}
}

\begin{document}

\maketitle              % typeset the title of the contribution

\begin{abstract}
$\lpmln$ is a probabilistic extension of answer set programs with the weight scheme adapted from Markov Logic. We study the concept of strong equivalence in $\lpmln$,  which is a useful mathematical tool for simplifying a part of an $\lpmln$ program without looking at the rest of it. We show that the verification of strong equivalence in $\lpmln$ can be reduced to equivalence checking in classical logic via a reduct and choice rules as well as to equivalence checking under the ``soft'' logic of here-and-there.  The result allows us to leverage an answer set solver for $\lpmln$ strong equivalence checking. The study also suggests us a few reformulations of the $\lpmln$ semantics using choice rules, the logic of here-and-there, and classical logic.
\end{abstract}
\section{Introduction}
$\lpmln$ is a probabilistic extension of answer set programs with the weight scheme adapted from Markov Logic \cite{richardson06markov}.  An $\lpmln$ program defines a probability distribution over all ``soft'' stable models, which do not necessarily satisfy all rules in the program, but the more rules with the bigger weights they satisfy, the bigger probabilities they get. 

The language turns out to be highly expressive to embed several other probabilistic logic languages, such as P-log \cite{baral09probabilistic}, ProbLog \cite{deraedt07problog}, Markov Logic, and Causal Models \cite{pearl00causality}, as described in  \cite{lee16weighted,balai16ontherelationship,lee17lpmln}. Inference engines for $\lpmln$, such as {\sc lpmln2asp}, {\sc lpmln2mln} \cite{lee17computing}, and {\sc lpmln-models} \cite{wang17parallel}, have been developed based on the reduction of $\lpmln$ to answer set programs and Markov Logic. 
$\lpmln$ is a basis of probabilistic action language $p{\cal BC}$+ \cite{lee18aprobabilistic}, which is defined as a high-level notation of $\lpmln$ to describe probabilistic transition systems.

As more results are built upon $\lpmln$, it becomes more critical to identify the equivalence between $\lpmln$ programs.  Similar to answer set programs,  $\lpmln$ programs $\mo{F}$ and $\mo{G}$ that have the same soft stable models with the same probability distribution are not necessarily equivalent in a stronger sense: when we add the same program $\mo{H}$ to each of $\mo{F}$ and $\mo{G}$, the resulting programs may have different soft stable models and different probability distributions. 

As in standard answer set programs, strong equivalence in $\lpmln$ is important in $\lpmln$ programming to simplify a part of an $\lpmln$ program without  looking at the rest of it and to verify the correctness of $\lpmln$ for the representation.  For instance, 
the following rules appearing in any program 
{
\[
\ba {rc}
   -(w_1 + w_2) :& a \lor b \\
   w_1 :& a \leftarrow b\\
   w_2 :&  b \leftarrow a
\ea 
\]
can be replaced by a simpler rule 
\[
\ba {rc}
   w_1 :&  a \\
   w_2 :&  b
\ea
\]}
without affecting the probability distribution over soft stable models.

However, because of the semantic differences, strong equivalence for answer set programs does not simply carry over to $\lpmln$.  First, the weights of rules play a role. Even for the same structure of rules, different assignments of weights make the programs no longer strongly equivalent. Also, due to the fact that soft stable models do not have to satisfy all rules, strongly equivalent answer set programs do not simply translate to strongly equivalent $\lpmln$ programs. 
For instance, 
$\{a\lor b, \ \ \bot\ar a, b\}$ is strongly equivalent to 
$\{a\ar \no\ b, \ \ b \ar \no\ a, \ \ \bot\ar a,b \}$, but its $\lpmln$ counterpart
$\{\alpha: a\lor b, \ \  \alpha:\bot\ar a, b\}$ is not strongly equivalent to 
$\{\alpha: a\ar \no\ b, \ \ \alpha: b \ar \no\ a, \ \ \alpha:\bot\ar a,b \}$: if we add 
$\{\alpha: a\ar b,\ \  \alpha: b\ar a\}$ to each of them, $\{a,b\}$ is a soft stable model of the former (by disregarding the rule $\alpha: \bot\ar a,b$) but not of the latter ({\bf c.f.} Example~\ref{ex:weak-strong}).

In this paper, we extend the notion of strong equivalence to $\lpmln$, and show that the verification of strong equivalence in $\lpmln$ can be reduced to equivalence checking in classical logic plus weight consideration. We also extend the logic of here-and-there to weighted rules, which provides a monotonic basis of checking $\lpmln$ strong equivalence. %We define soft equilibrium models and show that they are equivalent to soft stable models. 
The characterization of strong equivalence suggests us a few reformulations of the $\lpmln$ semantics using choice rules, the logic of here-and-there, and classical logic, which present us useful insights into the semantics.
%Based on the study, we present a method of checking strong equivalence using an answer set solver. 

The paper is organized as follows. After reviewing some preliminaries in Section~\ref{sec:prelim}, we present the definition of strong equivalence and some characterization of strong equivalence in terms of classical logic in Section~\ref{sec:se}. Then, we define the soft logic of here-and-there and soft equilibrium models, and show how soft logic of HT is related to strong equivalence in Section~\ref{sec:soft-ht}.  Then, we show another way to characterize strong equivalence in the style of second-order logic in Section~\ref{sec:se-delta} and use it to design a way to check strong equivalence using ASP solvers in Section~\ref{sec:se-solver}.

%-----------------------------------------------------------------------------------------
\section{Preliminaries}\label{sec:prelim}
%-----------------------------------------------------------------------------------------

%-----------------------------------------------------------------------------------------
\subsection{Review: Language $\lpmln$}\label{ssec:review-lpmln}
%-----------------------------------------------------------------------------------------
We first review the definition of a (deterministic) stable model for a propositional formula \cite{fer05}.
For any propositional formula $F$ and any set $X$ of atoms, the reduct $F^X$ is obtained from $F$ by replacing every maximal subformula of $F$ that is not satisfied by $X$ with $\bot$. 
Set $X$ is a {\em stable model} of $F$ if $X$ is a minimal model of the reduct $F^X$. 

We next review the definition of $\lpmln$ from~\cite{lee16weighted}.
 An $\lpmln$ program is a finite set of weighted formulas  $w: R$ where $R$ is a propositional formula \footnote{Same as in Markov Logic, we could allow schematic variables that range over the Herbrand Universe, and define the process of grounding accordingly. The result of this paper can be straightforwardly extended to that case.} and $w$ is a real number (in which case, the weighted rule is called {\em soft}) or $\alpha$ for denoting the infinite weight (in which case, the weighted rule is called {\em hard}). 
 %Throughout the paper, we assume that the language is propositional. Schematic variables can be introduced via grounding as standard in answer set programming. 
 
For any $\lpmln$ program $\mo{F}$ and any set $X$ of atoms,  
$\o{\mo{F}}$ denotes the set of usual (unweighted) formulas obtained from $\mo{F}$ by dropping the weights, and
$\mo{F}_X$ denotes the set of $w: R$ in $\mo{F}$ such that $X\models R$.

%In general, an $\lpmln$ program may even have stable models that violate some hard rules, which encode definite knowledge. However, throughout the paper, we restrict attention to $\lpmln$ programs whose stable models do not violate hard rules. 
%
%More precisely, 

Given an $\lpmln$ program $\mo{F}$, $\sm[\mo{F}]$ denotes the set of {\em soft stable models}: 
\[
\ba l
\{X\mid \text{$X$ is a (standard) stable model of $\o{\mo{F}_X}$}\}.
%\text{that satisfies all hard rules in $\Pi$} \} .%   $\o{\Pi^{\rm hard}}$}\}.
\ea
\]

By $\i{TW}(\mo{F})$ (``Total Weight" of $\mo{F}$) we denote the expression $exp({\sum\limits_{  w:R\in \mo{F}} w})$.
For any interpretation $X$, the weight of an interpretation $X$, denoted $W_{\mo{F}}(X)$, is defined as\footnote{We identify an interpretation with the set of atoms true in it.} 
{
\[
%\small
 W_{\mo{F}}(X) =
\begin{cases}
    \i{TW}(\mo{F}_X)  & 
%  exp\Bigg(\sum\limits_{w:R\;\in\; {\Pi}_I} w\Bigg) & 
      \text{if $X\in\sm[\mo{F}]$}; \\
  0 & \text{otherwise},
\end{cases}
\]
}
%where $\Pi^{\rm soft}$ consists of all soft rules in $\Pi$, 
and the probability of $X$, denoted $P_\mo{F}(X)$, is defined as
{%\small
\[
  P_\mo{F}(X)  = 
  \lim\limits_{\alpha\to\infty} \frac{W_\mo{F}(X)}{\sum\limits_{Y\in {\rm SM}[\mo{F}]}{W_\mo{F}(Y)}}. 
\]
}

Alternatively, the weight can be defined by counting the penalty of the interpretation \cite{lee17computing}.
More precisely, the penalty based weight of an interpretation $X$ is defined as the exponentiated negative sum of the weights of the rules that are not satisfied by $X$ (when $X$ is a stable model of $\overline{\mo{F}_X}$).  
Let 
\[
%\small
 W^{\rm pnt}_{\mo{F}}(X) =
\begin{cases}
   (TW(\mo{F}\setminus \mo{F}_X))^{-1} &
%  exp\Bigg(-\sum\limits_{w:R\;\in\; {\Pi} \text{ and } X\not\models R} w\Bigg) & 
      \text{if $X\in\sm[{\mo{F}}]$}; \\
  0 & \text{otherwise}, 
\end{cases}
\]
and
{%\small
$$ 
  P^{\rm pnt}_{\mo{F}}(X) = 
    \lim\limits_{\alpha\to\infty} \frac{W^{\rm pnt}_{\mo{F}}(X)}{\sum\limits_{Y\in {\rm SM}[{\mo{F}}]}{W^{\rm pnt}_{\mo{F}}(Y)}}.
$$ 
}

The following theorem tells us that the $\lpmln$ semantics can be reformulated using the concept of a penalty-based weight. 

\begin{thm}\label{thm:lpmln-pnt}\optional{thm:lpmln-pnt}
For any $\lpmln$ program $\mo{F}$ and any interpretation $X$, 
\[
\ba {rll}
%  W_{\mo{F}}(X) \propto W^{\rm pnt}_{\mo{F}}(X) % \times TW_{\Pi}
   W_{\mo{F}}(X) =& \i{TW}({\mo{F}})\times  W^{\rm pnt}_{\mo{F}}(X),  \\ % \times TW_{\Pi}
%\text{ \ \ \ \ and \ \ \ \ }
  P_{\mo{F}}(X) =& P_{\mo{F}}^{\rm pnt}(X).
\ea
\]
\end{thm}

%-----------------------------------------------------------------------------------------
\subsection{Review: Logic of Here and There}\label{ssec:review-ht}
%-----------------------------------------------------------------------------------------

Logic of here and there (Logic $HT$) is proven to be useful as a monotonic basis for checking strong equivalence \cite{lif01}, and equilibrium models \cite{pearce06equilibrium} are defined as a special class of minimal models in logic $HT$.  

An $HT$ interpretation is an ordered pair $\langle Y, X \rangle$ of sets of atoms such that $Y \subseteq X$, which describe ``two worlds": the atoms in $Y$ are true ``here" ($h$) and the atoms in $X$ are true ``there ($t$)." The worlds are ordered by $h < t$. 

For any $HT$ interpretation $\langle Y, X\rangle$, any world $w$, and any propositional formula $F$, we define when the triple $\langle Y,X, w\rangle$ satisfies $F$ recursively, as follows: 
%we write $\neg F$ for $F \rightarrow \bot$, $ \top$ for $\bot \rightarrow \bot$.
\begin{itemize}\addtolength{\itemsep}{0mm}
\item for any atom $F$, $\langle Y,X, h\rangle \modelsht F$ if $F \in Y$;  $\langle Y,X, t\rangle \modelsht F$ if $F \in X$.  
\item $\langle Y,X, w\rangle \not\modelsht \bot$.
\item $\langle Y,X, w\rangle \modelsht F \land G$ if %\vspace{-1mm}
%\begin{center}
$\langle Y,X, w\rangle \modelsht F$ and $\langle Y,X, w\rangle \modelsht G$.
%\end{center}
\item $\langle Y,X, w\rangle \modelsht F \lor G$ if %\vspace{-1mm}
%\begin{center}
$\langle Y,X, w\rangle \modelsht F$ or $\langle Y,X, w\rangle \modelsht G$.
%\end{center}
\item $\langle Y,X, w\rangle \modelsht F \rightarrow G$ if for every world such that $w \leq w' $, 
%\vspace{-1mm}
%\begin{center}
$\langle Y,X, w'\rangle \not\modelsht F$ or $\langle Y,X, w'\rangle \modelsht G$. 
%\end{center}
\end{itemize}

\begin{definition}\label{def:ht-models}
We say that an $HT$ interpretation $\langle Y, X\rangle$ satisfies $F$ (symbolically, $\langle Y, X\rangle\modelsht F$) 
if $\langle Y,X, h\rangle$ satisfies $F$. An $HT$ model of $F$ is an $HT$ interpretation that satisfies $F$. 
\end{definition}
%If $\langle Y,X, H\rangle$ satisfies $F$, then $\langle Y,X, t\rangle$ satisfies $F$ by induction. 

Equilibrium models are defined as a special class of minimal models in logic $HT$ as follows.

\begin{definition}\label{def:equil}
An $HT$ interpretation $\langle Y,X\rangle$ is {\em total} if $Y = X$. A total $HT$ interpretation $\langle X, X\rangle$ is an {\em equilibrium model} of a propositional formula $F$ if 
\begin{itemize}
\item $\langle X, X\rangle \modelsht F$, and
\item for any proper subset $Y$ of $X$, 
          $\langle Y,X\rangle\not\modelsht F$.
\end{itemize}
\end{definition}

A natural deduction system for logic $HT$ can be obtained from the natural deduction system for classical logic by dropping the law of excluded middle 
$
F \lor \neg F
$
from the list of deduction rules 
and by adding
%A formalization of $HT$ can be obtained from intuitionistic logic by adding 
the axiom schema
$
F \lor (F \rightarrow G) \lor \neg G. 
$
From the deduction system, we can derive the weak law of excluded middle
$\neg F \lor \neg\neg F$.

Theorem~1 from \cite{lif01} shows that strong equivalence between two answer set programs coincides with equivalence in logic $HT$.  The deduction rules above can be used for checking strong equivalence. 

%-----------------------------------------------------------------------------------------
\section{Strong Equivalence in $\lpmln$}\label{sec:se}
%-----------------------------------------------------------------------------------------

We define the notions of weak and strong equivalences, naturally extended from those for the standard stable model semantics. 

\begin{definition} \label{def:we}
$\lpmln$ programs $\mo{F}$ and $\mo{G}$ are called {\em weakly equivalent} to each other if 
\[
   P_{\mo{F}}(X) = P_{\mo{G}}(X)
\]
for all interpretations $X$. 
\end{definition}

\begin{definition} \label{def:se}
$\lpmln$ programs $\mo{F}$ and $\mo{G}$ are called {\em strongly equivalent} to each other if, for any $\lpmln$ program $\mo{H}$,
\[
   P_{\mo{F}\cup \mo{H}}(X) = P_{\mo{G}\cup \mo{H}}(X)
\]
for all interpretations $X$. 
\end{definition}

Note that strong equivalence implies weak equivalence, but not vice versa. 

\begin{example} \label{ex:weak-strong}
Consider two programs $\mo{F}$ and $\mo{G}$ \footnote{We identify $F\ar G$ with $G\rar F$ and $\ar F$ with $F\rar\bot$.}
\[
\ba {lrcllrcl}
\mo{F} & 2: & a \lor b   &  \hspace{5mm} &   \mo{G} & 1: & a \leftarrow \neg b \\
             & 1: & \leftarrow a\land b &   &                & 1: & b \leftarrow \neg a\\
             &     &                                 &   &      &  1: & \leftarrow a\land b. \\
\ea 
\] 
The programs are weakly equivalent, but not strongly equivalent.  One can check their probability distributions over soft stabel models are identical. 
However, for 
$$\mo{H} = \{1:\ \ a \leftarrow b,\ \ \  1:\ \ b\leftarrow a\},$$ 
set $\{a, b\}$ is a soft stable model of $\mo{F \cup H}$ but not of $\mo{G \cup H}$, so that $P_{\mo{F\cup H}}(\{a, b\})$ is $e^4/Z$ ($Z$ is a normalization factor) but $P_{\mo{G \cup H}}(\{a, b\})$ is 0.
\end{example}

We call an expression of the form $e^{c_1+c_2\alpha}$, where $c_1$ is a real number accounting for the weight of soft rules and $c_2$ is an integer accounting for the weight of hard rules, a {\em w-expression}. Then Definition~\ref{def:se} can be equivalently rewritten as follows: $\mo{F}$ and $\mo{G}$ are strongly equivalent to each other if
there is a $w$-expression $c$ such that for any $\lpmln$ program $\mo{H}$,  
\beq
%   W_{\mo{F}\cup \mo{H}}(X) = e^{c_1+c_2\alpha}\times W_{\mo{G}\cup \mo{H}}(X).
   W_{\mo{F}\cup \mo{H}}(X) = c\times W_{\mo{G}\cup \mo{H}}(X)
\eeq{w-se}
for all interpretations $X$. 
The w-expression $c$ accounts for the fact that the weights are ``proportional" to each other, so the probability distribution remains the same.

In view of Theorem~\ref{thm:lpmln-pnt}, it is also possible to use the penalty based weights, i.e., the equation 
\[
   W^{\rm pnt}_{\mo{F}\cup \mo{H}}(X) = c\times W^{\rm pnt}_{\mo{G}\cup \mo{H}}(X)
\]
can be used in place of~\eqref{w-se}. 

By definition, every interpretation that has a non-zero weight is a soft stable model. Thus Definition~\ref{def:se} implies that the $\lpmln$ programs are ``structurally equivalent" to each other, which is defined as follows.

%for strongly equivalent $\lpmln$ programs $\mo{F}$ and $\mo{G}$, for any $\lpmln$ programs $\mo{H}$, programs $\mo{F}\cup\mo{H}$ and $\mo{G}\cup\mo{H}$ have the same set of soft stable models.
 
\begin{definition} \label{def:struct-equiv}
$\lpmln$ programs $\mo{F}$ and $\mo{G}$ are {\em structurally  equivalent} if, for any $\lpmln$ program $\mo{H}$, programs $\mo{F}\cup\mo{H}$ and $\mo{G}\cup\mo{H}$ have the same set of soft stable models.
\end{definition}

Strong equivalence implies structural equivalence, but not vice versa.

{\begin{prop} \label{prop:se-str}
If $\lpmln$ programs $\mo{F}$ and $\mo{G}$ are strongly equivalent, then 
they are structurally equivalent as well. 
\end{prop}}

The fact that $\lpmln$ programs $\mo{F}$ and $\mo{G}$ are structurally equivalent does not follow from the fact that ASP programs $\o{\mo{F}}$ and $\o{\mo{G}}$ are strongly equivalent. 

\medskip\noindent
{\bf Example~\ref{ex:weak-strong} Continued}\ \ 
{\sl 
In Example~\ref{ex:weak-strong}, two ASP programs $\o{\mo{F}}$ and $\o{\mo{G}}$ are strongly equivalent (in the sense of standard answer set programs) but 
${\mo{F}}$ and ${\mo{G}}$ are not structurally equivalent, and consequently not strongly equivalent. If we add $\mo{H} = \{1: a \ar b, \ \ 1: b\ar a\}$ to each program, $X=\{a,b\}$ is a soft stable model of $\mo{F}\cup\mo{H}$ but not of
$\mo{G}\cup\mo{H}$.
}\medskip

%Let $TW(F)$ be $\sum\limits_{  w:R\in F} w$.

The following theorem shows a characterization of strong equivalence that does not need to consider adding all possible $\lpmln$ programs $\mo{H}$.  Similar  to Proposition~2 from \cite{fer05}, it shows that the verification of strong equivalence in $\lpmln$ can be reduced to equivalence checking in classical logic plus weight checking. 

\begin{thm} \label{thm:se-char} 
For any $\lpmln$ programs $\mo{F}$ and $\mo{G}$, program $\mo{F}$ is strongly equivalent to $\mo{G}$ if and only if there is a w-expression $c$ such that for every interpretation $X$, 
\begin{enumerate}
\item $TW(\mo{F}_X) = c \times TW(\mo{G}_X)$, and 
%\item (or equivalently, $P(\mo{F}_X) = c \times TW(\mo{G}_X)$, and 
\item $(\o{\mo{F}_X})^X$ and  $(\o{\mo{G}_X})^X$ are classically equivalent. 
\end{enumerate}
\end{thm}

Recall that $TW({\mo{F}}_X)$ is simply an exponentiated sum of the weights of the rules that are true in $X$.  Condition~1 of Theorem~\ref{thm:se-char} does not require to check whether $X$ is a soft stable model or not.  \footnote{Instead, Condition~2 ensures that they are structurally equivalent as shown in {\bf Theorem on Soft Stable Models} below.} 
In view of Theorem~\ref{thm:lpmln-pnt}, 
%one could also equivalently add up the weights of rules that are not true in $X$ and exponentiate the sum, i.e., 
the condition %Condition~1 of Theorem~\ref{thm:se-char} 
can be replaced with 
\begin{enumerate}
\item[{\sl 1'.}] $TW(\mo{F}\setminus\mo{F}_X) = c \times TW(\mo{G}\setminus\mo{G}_X)$.
\end{enumerate}
without affecting the correctness of Theorem~\ref{thm:se-char}.

%$TW({\mo{F}}\setminus\mo{F}_X)$, which consider rules that are not true in $X$.
%Again, in Condition~1 of Theorem~\ref{thm:se-char} could refer to the penalty based weights. 

\begin{example}\label{ex:main}
Consider two programs 
\[
\ba {lrcllrcl}
\mo{F} ~ & 0: & \neg a  & \hspace{5mm} & \mo{G} ~ & 2: & \neg a \lor b \\
            & 2: & b\ar a  &                         &              &1: &  a\lor \neg a  \\
            & 3: & a \ar \neg \neg a \\
\ea 
\] 
The programs are strongly equivalent to each other.
The following table shows that Conditions 1,2 of  Theorem~\ref{thm:se-char} are true  in accordance with the theorem. 

%two strongly equivalent programs $\mo{F}$ and $\mo{G}$.
%Here is a positive example that $F$ and $G$ are strong equivalent in which $(F_X)^X \leftrightarrow (G_X)^X$.

\begin{table}[h!] 
\begin {center}
\begin{tabular}{c|c|c|c|c}
%    & $\mo{F}$ & $\mo{G}$} \\
%\empty & $0: \neg a $ & $2: \neg a \lor b $ \\
%\empty & $3: a \leftarrow \neg \neg a$ & $1: a \leftarrow \neg \neg a$ \\ 
%\empty & $2: b \leftarrow a$ & \empty \\
%\hline\hline
$X$  & $TW(\mo{F}_X)$ & $TW(\mo{G}_X)$ & $(\o{\mo{F}_X})^X$ & $(\o{\mo{G}_X})^X$ \\ \hline \hline
$\phi$ &$e^{5}$ & $e^{3}$ & $\top$ & $\top$ \\
\hline
$\{a\}$ &$e^3$ & $e^1$ & $a$ & $a$ \\
\hline
$\{b\}$ & $e^5$ & $e^3$ & $\top$ & $\top$ \\
\hline
$\{a, b\}$& $e^5$ & $e^3$  & $a\land b$ & $a\land b$
\end{tabular}
\end{center}
%\vspace{-3mm}
\caption{$(\o{\mo{F}_X})^{X}$ and $(\o{\mo{G}_X})^{X}$}
%\vspace{-2mm}
\label{tab:table1}
\end{table}
%\FloatBarrier

%[[ Man: please complete the table ]]

Note that $TW({\mo{F}_X}) = e^2 \times TW(\mo{G}_X)$. However, if we replace rule \ \ $3:\ \ a \ar \neg\neg a$\ \  in $\mo{F}$ with \ \ $3:\ \ a \ar a$\ \  to result in $\mo{F}'$, then $\mo{F}'$ and $\mo{G}$ are not strongly equivalent: for 
$$
\mo{H}=\{1 :a \leftarrow b, \ \ 1: b \leftarrow a\}
$$ 
$\{a, b\}$ is a soft stable model of $\mo{G}\cup\mo{H}$ with the weight $e^5$, but it is not a soft stable model $\mo{F'}\cup\mo{H}$, so its weight is $0$. 
In accordance with Theorem~\ref{thm:se-char}, 
$(\o{\mo{F'}_{\{a, b\}}})^{\{a, b\}}$ is
not equivalent to 
$(\o{\mo{G}_{\{a, b\}}})^{\{a, b\}}$. 
The former is equivalent to $\{  b \leftarrow a\}$, 
and the latter is equivalent to 
$\{a \land b\}$.

Even if the programs have the same soft stable models, the different weight assignments may make them not strongly equivalent. For instance, replacing the first rule in $\mo{G}$ by \hbox{$3: \neg a \lor b$} to result in $\mo{G}'$, we have $TW(\mo{F}_{\phi}) = e^1 \times TW(\mo{G'}_{\phi})$ and $TW(\mo{F}_{\{a\}}) = e^2 \times TW(\mo{G'}_{\{a\}})$, so there is no single w-expression $c$ such that $TW(\mo{F}_X) = c \times TW(\mo{G'}_X)$.
% so they are no longer strong equivalent.
%{\cmag  [[ explain; weights]] }
\end{example}

Choice rules are useful constructs in answer set programming, and they turn out to have an interesting position in the semantics of $\lpmln$. We consider a general form of choice rules that is not limited to atoms. For any propositional formula $F$, by $\{F\}^{\rm ch}$ we denote the formula $F\lor\neg F$. 
The following proposition tells us that choice rules can be alternatively represented in $\lpmln$ with the weight $0$ rule.

\begin{prop} \label{prop:1}
For any formula $F$, 
the weighted formula $0:F$ is strongly equivalent to $w:\ \{F\}^{\rm ch}$, where $w$ is any real number or $\alpha$.
\end{prop}

%\begin{prop}
%Let $\mo{H}$ be a weighted formula $w:H$ that is strongly equivalent to $w: \top$ ($w$ is a real or $\alpha$). For any $\lpmln$ program $\mo{F}$, program~$\mo{F}\cup\mo{H}$ is strongly equivalent to $\mo{F}$.
%\end{prop}
%

%-----
The following fact can also be useful for simplification.

\begin{prop}\label{prop:2}
Let $\mo{H}$ be an $\lpmln$ program that is structurally equivalent to $w: \top$ or $w:\bot$ ($w$ is a real number or $\alpha$). For any $\lpmln$ program $\mo{F}$, program~$\mo{F}\cup\mo{H}$ is strongly equivalent to $\mo{F}$.
\end{prop}
For example, adding $\mo{H} = \{w_1: a \land \neg a,\ \  w_2: a \leftarrow  a \}$ to $\mo{F}$, one can easily see $\mo{F}$ and $\mo{F}\cup\mo{H}$ are strongly equivalent.

{Interestingly, some facts about strong equivalence known in answer set programs do not simply carry over to $\lpmln$ strong equivalence. The fact that, for any propositional formulas $F$,$G$, and $K$,
\[
 (F \rar G) \rar K 
\] 
is strongly equivalent to 
\[
\ba l
 (G \lor \neg F) \rar K \\
  K \lor F \lor \neg G
\ea 
\]
is a key lemma to prove that any propositional formulas can be turned into the logic program syntax \cite{caba07}. The result is significant because it allows stable models of general syntax of formulas to be computed by converting into rule forms and computed by standard answer set solvers, as done in system {\sc f2lp}. However, it turns out that the transformation does not work under $\lpmln$, i.e., there are some formulas $F$, $G$, $K$ such that 
\beq
\ba {rl}
w:& (F \rar G) \rar K 
\ea
\eeq{f2lp1}
is not strongly equivalent to 
\beq
\ba {rl}
w_1: & (G \lor \neg F) \rar K \\
w_2: & K \lor F \lor \neg G
\ea 
\eeq{f2lp2}
regardless of weights $w$, $w_1$, $w_2$.
For example, assuming $F$, $G$, $K$ are atoms, and take interpretation $X = \{F, G\}$.
\[
\ba {rcl}
(((F \rightarrow G) \rightarrow K)_X)^{X}
    &\Leftrightarrow& \bot  \\
((\{(G \lor \neg F) \rightarrow K,\ \ K \lor F \lor \neg G\})_{X})^{X} &\Leftrightarrow& F^{X}.
\ea 
\]
So Condition~2 of Theorem~\ref{thm:se-char} does not hold, and it follows that \eqref{f2lp1} is not strongly equivalent to \eqref{f2lp2}.
}
\footnote{
Of course, \eqref{f2lp1} is strongly equivalent to 
\[
\ba {rl}
w: &  ((G \lor \neg F) \rar K)\land ( K \lor F \lor \neg G) 
\ea
\]
but the latter is not in a rule form.}

%-----------------------------------------------------------------------------------------
\subsection{Reformulation of $\lpmln$ Using Choice Rules} \label{ssec:choice}
%-----------------------------------------------------------------------------------------
The second condition of Theorem~\ref{thm:se-char} is equivalent to the fact that for any $\lpmln$ program $\mo{H}$, programs $\mo{F}\cup\mo{H}$ and $\mo{G}\cup\mo{H}$ have the same soft stable models. Throughout the paper, we show that the condition can be represented in several different ways. We start with the following version that uses choice rules.

%In ASP, the choice rules are applied to a set of atoms only. Here, we apply to any propositional formulas. For any propositional formula $F$, $\{F\}^{\rm ch}$ denotes the formula $F\lor\neg F$. 
We extend the notion of choice rules to a set of formulas as follows: for a set $\Gamma$ of propositional formulas, $\{\Gamma\}^{\rm ch}$ denotes the set of choice formulas $\{ \{F\}^{\rm ch} \mid F \in \Gamma \}$.

\medskip\noindent
{\bf Theorem on Soft Stable Models}\ \ \ 
{\sl 
For any $\lpmln$ program $\mo{F}$ and $\mo{G}$,
%and for any sets $Y$, $X$ of atoms such that $Y\subseteq X$, 
the following conditions are equivalent. 
{
\begin{enumerate}[(a)]
\item $\mo{F}$ and $\mo{G}$ are structurally equivalent. 
% For any $\lpmln$ program $\mo{H}$, programs $\mo{F}\cup\mo{H}$ and $\mo{G}\cup\mo{H}$ have the same soft stable models. 

%\item For any set $X$ of atoms, $\o{\mo{F}_X}$ and $\o{\mo{G}_X}$ are strongly equivalent.

\item For any set $X$ of atoms, $(\o{\mo{F}_X})^X$ and $(\o{\mo{G}_X})^X$ are classically equivalent.

\item For any set $X$ of atoms, $(\{\o{\mo{F}}\}^{\rm ch})^X$ and $(\{\o{\mo{G}}\}^{\rm ch})^X$ are classically equivalent.

\end{enumerate}
}
}\medskip

Thus, Theorem \ref{thm:se-char} remains valid if we replace Condition~2 in it with 
%\smallskip
\begin{enumerate}
\item[{\sl 2$'$.}] $(\{\o{\mo{F}}\}^{\rm ch})^X$ and $(\{\o{\mo{G}}\}^{\rm ch})^X$ are classically equivalent.
\end{enumerate}
\smallskip 

As a side remark, {\bf Theorem on Soft Stable Models} also tells us an equivalent characterization of soft stable models, which in turn leads to a reformulation of $\lpmln$ semantics. 

\begin{prop} \label{prop:3}
For any $\lpmln$ program $\mo{F}$, $X$ is a soft stable model of $\mo{F}$ iff $X$ is a (standard) stable model of~$\{\o{\mo{F}}\}^{\rm ch}$.
\end{prop}

%-----------------------------------------------------------------------------------------
\section{Soft Logic of Here and There}\label{sec:soft-ht}
%-----------------------------------------------------------------------------------------

We extend the logic of here-and-there and the concept of equilibrium models to $\lpmln$ programs as follows. 

\begin{definition}\label{def:soft-ht-models}
An $HT$ interpretation $\langle Y, X\rangle$ is called a {\em soft $HT$ model} of an $\lpmln$ program $\mo{F}$ if, for every rule $w:R$ in $\mo{F}_{X}$, $\langle Y, X\rangle$ satisfies $R$. In other words, $\langle Y, X\rangle$ is a soft $HT$ model of $\mo{F}$ iff 
$\langle Y, X\rangle$ is an $HT$ model of $\o{\mo{F}_X}$.
\end{definition}

%[[ Note that by definition of $\mo{F}_X$, $X$ satisfies all rules there, so we do not need to repeat.]]

%[[ We do not need the first condition ]]

%[[ The weight of the equilibrium models are defined in the same way as in stable models in Section XX.]]

We extend the {\bf Theorem on Soft Stable Models} to consider HT models as follows. We omit repeating conditions (b), (c). 

\medskip\noindent
{\bf Theorem on Soft Stable Models}\ \ \ 
{\sl 
For any $\lpmln$ program $\mo{F}$ and $\mo{G}$,
%and for any sets $Y$, $X$ of atoms such that $Y\subseteq X$, 
the following conditions are equivalent. 
\begin{enumerate}[(a)]
{ \item $\mo{F}$ and $\mo{G}$ are structurally equivalent.  }
%
%\item For any set $X$ of atoms, $\o{\mo{F}_X}$ and $\o{\mo{G}_X}$ are strongly equivalent.
%
%\item For any set $X$ of atoms, $(\o{\mo{F}_X})^X$ and $(\o{\mo{G}_X})^X$ are classically equivalent.
%
%\item For any set $X$ of atoms, $(\{\o{\mo{F}}\}^{\rm ch})^X$ and $(\{\o{\mo{G}}\}^{\rm ch})^X$ are classically equivalent.

\item[(d)] $\mo{F}$ and $\mo{G}$ have the same set of soft $HT$ models.

\item[(e)]  For any set $X$ of atoms, $\o{\mo{F}_X} \lrar \o{\mo{G}_X}$ is provable in logic $HT$. 

%\item[(f)]  For any set $X$ of atoms, $\o{\mo{F}_X}$ and $\o{\mo{G}_X}$ have same set of $HT$ models.

\item[(f)]  $(\{\o{\mo{F}}\}^{\rm ch}) \lrar (\{\o{\mo{G}}\}^{\rm ch})$ is provable in logic $HT$. 

%\item[(h)]   $(\{\o{\mo{F}}\}^{\rm ch})$ and $(\{\o{\mo{G}}\}^{\rm ch})$ have same set of HT models. 

\end{enumerate}
}\medskip

Again, any of the conditions $(d)$, $(e)$, $(f)$ can replace Condition 2 of  Theorem~\ref{thm:se-char} without affecting the correctness.

{
\medskip\noindent
{\bf Example~\ref{ex:main} Continued}\ \ \ 
{\sl We consider soft $HT$ models of $\mo{F}$, $\mo{G}$ and $\mo{F}'$ in Example~\ref{ex:main}.

\begin{table}[h!]
{%\footnotesize
\begin {center}
\begin{tabular}{c|c|c|c}
  
% \textbf{$X$} & \textbf{$F$} & \textbf{$G$} &  $\mo{F}'$\\
% \empty & $\{\neg a\}^{\rm ch} $ & $ \{\neg a \lor b\}^{\rm ch} $ \\
% \empty & $\{\neg\neg a\rar a\}^{\rm ch}$ & $\{\neg\neg a\rar a\}^{\rm ch}$ \\
% \empty & $\{a\rar b\}^{\rm ch}$ & \empty \\
% \hline
$X$ & $\mo{F}$  & $\mo{G}$  &  $\mo{F'}$ \\ \hline
$\langle \phi, \phi \rangle $ & Yes & Yes & Yes\\
\hline
$\langle \phi, \{a\}\rangle $ & No & No & Yes\\
\hline
$\langle \{a\}, \{a\}\rangle $ & Yes & Yes & Yes\\
\hline
$\langle \phi, \{b\}\rangle $ & Yes & Yes & Yes\\
\hline
$\langle \{b\}, \{b\}\rangle $ & Yes & Yes & Yes\\
\hline
$\langle \phi, \{a, b\}\rangle $ & No & No & Yes\\
\hline
$\langle \{a\}, \{a, b\}\rangle $ & No & No & No\\
\hline
$\langle \{b\}, \{a, b\}\rangle $ & No & No & Yes\\
\hline
$\langle \{a, b\}, \{a, b\}\rangle $& Yes & Yes & Yes\\
\end{tabular}
\end{center}
}
%\vspace{-2mm}
\caption{Soft $HT$ models of $\mo{F}$, $\mo{G}$, and $\mo{F'}$}
%\vspace{-3mm}
\label{tab:table3}

\end{table}
%\FloatBarrier

From Table~\ref{tab:table3}, we see that $\mo{F}$ and $\mo{G}$ have the same set of soft $HT$ models.
%is equivalent to saying that they are provable in logic $HT$. 
%Also, it  illustrates the equivalence between theorem $g$ and theorem 2. 

%\end{example}
}\medskip
}

Condition~(f) allows us to prove the structural equivalence between two $\lpmln$ programs by using deduction rules in logic $HT$.
\begin{example}
% to show that we can check whether two $\lpmln{}$ programs $\mo{F}$ and $\mo{G}$ are strongly equivalent by checking whether $ \{\mo{\o F}\}^{\rm ch} \leftrightarrow \{\mo{\o G}\}^{\rm ch}$ is provable in $HT$ logic. 
Consider $\lpmln$ programs $\mo{F}$ and $\mo{G}$:
\[
\ba {lrcllrcl}
\mo{F} ~ & 2: & \neg a \lor b   &  \hspace{5mm} &   \mo{G} ~ & 2: \neg \neg a \rightarrow b \\
\ea 
\] 

We check that $ \{\mo{\o F}\}^{\rm ch} \leftrightarrow \{\mo{\o G}\}^{\rm ch}$ is provable in logic $HT$.
Recall that 

\[
\ba l
\{\mo{\o F}\}^{\rm ch}= (\neg a\lor b) \lor \neg (\neg a\lor b) \\
\{\mo{\o G}\}^{\rm ch}= (\neg \neg a \rightarrow b) \lor \neg (\neg \neg a \rightarrow b).
\ea
\]

\noindent
{\bf Left-to-right:}
Assume $(\neg a\lor b) \lor \neg (\neg a\lor b)$.

\noindent
Case 1: Assume $(\neg a\lor b)$. Then $\neg \neg a \rightarrow b$ is  intuitionistically derivable, so derivable in logic $HT$ as well. 

\noindent
Case 2: Assume $\neg (\neg a\lor b)$. Then $ \neg( \neg \neg a \rightarrow b)$ is  intuitionistically derivable {(Glivenko's Theorem)}.

\medskip\noindent
{\bf Right-to-left:}
Assume $(\neg \neg a \rightarrow b) \lor \neg (\neg \neg a \rightarrow b)$.

\noindent
Case 1: Assume $\neg \neg a \rightarrow b$. Then $\neg a \lor b$ can be derived from the weak law of excluded middle $\neg a \lor \neg \neg a$. 
%\begin{enumerate}

\noindent
Case 2: Assume $ \neg (\neg \neg a \rightarrow b)$. Then $\neg(\neg a \lor b)$ is intuitionistically derivable {(Glivenko's Theorem)}.

\medskip
In view of the equivalence between Conditions (a) and (f) of {\bf Theorem on Soft Stable Models}, we conclude that $\mo{F}$ and $\mo{G}$ are structurally equivalent.
\end{example}
%[[ This gives us a way to use deduction : example should use deduction ]]

%[[ Proof theoretic account of STRONG EQUIVALENCE ]]

%-----------------------------------------------------------------------------------------
\subsection{Soft Equilibrium Models}\label{ssec:soft-equil}
%-----------------------------------------------------------------------------------------

\begin{definition}\label{def:soft-equil}
A soft $HT$ interpretation is called {\em total} if $Y = X$. A total soft $HT$ interpretation $\langle X, X\rangle$ is a {\em soft equilibrium model} of an $\lpmln$ program $\mo{F}$ if, for any proper subset $Y$ of $X$, 
              $\langle Y, X\rangle$ is not a soft $HT$ model of  $\mo{F}$.
%\begin{itemize}
%%\item $\langle X, X\rangle$ is a soft $HT$ model of $\mo{F}$, and 
%%\item  for any proper subset $Y$ of $X$, 
%%              $\langle Y, X\rangle$ is not a soft $HT$ model of  $\mo{F}$.
%\end{itemize} 
\end{definition}

In comparison with Definition~\ref{def:equil}, Definition~\ref{def:soft-equil} omits the condition that $\langle X,X\rangle$ satisfies $\o{\mo{F}_X}$ because the condition is trivially satisfied by the definition of $\mo{F}_X$. 

The following lemma tells us how soft HT models are related to the reducts in $\lpmln$.

{
\begin{lemma} \label{lem:main}
For any $\lpmln$ program $\mo{F}$ and any sets $Y,X$ of atoms such that $Y\subseteq X$, the following conditions are equivalent:
\begin{enumerate}[(a)]
    \item $\langle Y, X\rangle$ is a soft $HT$ model of $\mo{F}$.
    {\item $Y$ satisfies $(\o{\mo{F}_{X}})^{X}$.
    \item $Y$ satisfies $(\{\o{\mo{F}}\}^{\rm ch})^X$.}
\end{enumerate}
%$\langle Y, X\rangle$ is a soft $HT$ model of $\mo{F}$ iff $Y$ is a model of $(\o{\mo{F}_{X}})^{X}$ in the sense of classical logic.
\end{lemma}
}

% \begin{prop}
% $\langle Y, X\rangle  $ is a soft $HT$ model of $\lpmln$ program $P$ iff $Y \models (P_{X})^{X}$.
% \end{prop}

% By lemma \ref{lemma_reduct}, $\langle Y, X\rangle  $ is a soft $HT$ model of $P$ iff $\langle Y, X\rangle  $ is a soft $HT$ model of $P^{X}$. By lemma \ref{lemma_noneg},  $\langle Y, X\rangle  $ is a soft $HT$ model of $P^{X}$ iff $Y \models (P_{X})^{X}$. 

From the lemma, we conclude:
\begin{prop} \label{prop:4}
A set $X$ of atoms is a soft stable model of $\mo{F}$ iff {$\langle X, X\rangle$} is a soft equilibrium model of $\mo{F}$.
\end{prop}

\medskip\noindent
{\bf Example~\ref{ex:main} Continued}\ \ 
{\sl
Table \ref{tab:table1} shows that $\mo{F}$ and $\mo{G}$ have three soft stable models, which are $\phi, \ \{a\}, \ \{a, b\}$. Table \ref{tab:table3} shows that $\mo{F}$ and $\mo{G}$ have three equilibrium models, which are $\langle \phi, \phi \rangle, \ \langle \{a\}\, \{a\} \rangle, \ \langle \{a, b\},  \{a, b\} \rangle$. On the other hand, $\mo{F'}$ has   only one equilibrium model, $\langle \phi, \phi \rangle$, which provides another account for the fact that $\mo{F'}$ and $\mo{G}$ have different soft stable models.
}\medskip 

The weight of a soft equilibrium model can be defined the same as the weight of a soft stable model as defined in Section~\ref{ssec:review-lpmln}.

%-----------------------------------------------------------------------------------------
\section{Strong Equivalence by Reduction to Classical Logic}  \label{sec:se-delta}
%-----------------------------------------------------------------------------------------

We extend the theorem on stable models as follows. Let ${\bf p}$ be the propositional signature.
Let ${\bf p}'$ be the set of new atoms $p'$ where $p\in {\bf p}$.
For any formula $F$, $\Delta_{{\bf p}'}(F)$ is defined recursively:
\begin{itemize}
\item $\Delta_{{\bf p}'}(p) = p'$ for any atomic formula $p\in {\bf p}$;
%\item $\Delta_{{\bf p}'}(F) = \{F'\}^{\rm ch}$ for any atomic formula $F$;
\item $\Delta_{{\bf p}'}(\neg F) = \neg F$;
\item $\Delta_{{\bf p}'}(F \land G) = \Delta_{{\bf p}'}(F) \land \Delta_{{\bf p}'}(G)$;
\item $\Delta_{{\bf p}'}(F \lor G) = \Delta_{{\bf p}'}(F) \lor \Delta_{{\bf p}'}(G)$;
\item $\Delta_{{\bf p}'}(F \rightarrow G) = (\Delta_{{\bf p}'}(F) \rightarrow \Delta_{{\bf p}'}(G)) \land (F \rightarrow G)$.
\end{itemize}

\medskip
Lemma~\ref{lem:main} is extended to $\Delta$ as follows. 

\medskip\noindent
{\bf Lemma~\ref{lem:main}'} \ \ \
{\sl 
Let $X, Y\subseteq {\bf p}$ and $Y' = \{p'\in {\bf p}' \mid p\in Y\}$.
Each of the following conditions is equivalent to each of Conditions~(a),(b),(c) of Lemma~\ref{lem:main}. 

\begin{enumerate}% \addtolength{\itemsep}{
%\item $\langle Y,X\rangle$ is a soft $HT$-model of $\mo{F}$.
%\item $\langle Y,X\rangle$ is a $HT$-model of $\overline{\mo{F}_X}$.
%\item $Y\models (\o{\mo{F}_X})^X$.
%\item $Y\models (\{\o{\mo{F}}\}^{\rm ch})^X$.
\item[(d)] $Y'\cup X$ satisfies $\Delta_{{\bf p'}}(\o{\mo{F}_X})$.
\item[(e)] $Y'\cup X$ satisfies $\Delta_{{\bf p'}}(\{\o{\mo{F}}\}^{\rm ch})$.
\end{enumerate}
}

\medskip\noindent
{\bf Theorem on Soft Stable Models}\ \ \ 
{\sl 
For any $\lpmln$ programs $\mo{F}$ and $\mo{G}$,
%and for any sets $Y$, $X$ of atoms such that $Y\subseteq X$, 
the following conditions are equivalent. 
\begin{enumerate}[(a)]
\item  $\mo{F}$ and $\mo{G}$ are structurally equivalent.

\item[(g)]  For any set $X$ of atoms, $\{p'\rar p \mid p \in {\bf p}\}$ 
entails $\Delta_{{\bf p'}}(\o{\mo{F}_X})\lrar \Delta_{{\bf p'}}(\o{\mo{G}_X})$
 (in the sense of classical logic).

\item[(h)]
$\{p'\rar p \mid p \in {\bf p}\}$ entails
   $ \Delta_{{\bf p'}}(\{\o{\mo{F}}\}^{\rm ch})\lrar\Delta_{{\bf p'}}(\{\o{\mo{G}}\}^{\rm ch})$
(in the sense of classical logic).
\end{enumerate}
}\medskip

The equivalence between Conditions (a) and (h)  of {\bf Theorem on Soft Stable Models} tells us the structural equivalence checking reduces to satisfiability checking. It also indicates the structural equivalence checking between $\lpmln$ programs is no harder than checking strong equivalence between standard answer set programs. In conjunction with Condition~1 of Theorem~\ref{thm:se-char},  the complexity of $\lpmln$ strong equivalence checking is no harder than checking strong equivalence for standard answer set programs.

{
\begin{thm} \label{thm:complexity}
The problem of determining if two $\lpmln$ programs are strongly equivalent is co-NP-complete.
\end{thm}
}

%-----------------------------------------------------------------------------------------
\subsection{Reformulation  of $\lpmln$ in Classical Logic}  \label{ssec:reform-delta}
%-----------------------------------------------------------------------------------------

The following proposition relates $\Delta$ to soft stable models. 

\begin{prop} \label{prop:delta}
%Let $Y\subseteq X\subseteq {\bf p}$, and $Y' = \{p'\in {\bf p}' \mid p\in Y\}$.
For any $\lpmln$ program $\mo{F}$, set $X$ is a soft stable model of $\mo{F}$ iff there is no strict subset $Y$ of $X$ such that $Y'\cup X$ satisfies $\Delta_{\bf p'}(\{\o{\mo{F}}\}^{\rm ch})$.
%\[
%   \Big(\bigwedge_{p'\in {\bf p}', p\in {\bf p}} (p'\rar p)\Big) \rar \Delta(\{\o{\mo{F}}\}^{\rm ch}).
%\]
%where ${\bf p}$ is a set of atoms occurring in $\mo{F}$ and
%${\bf p}'$ is $\{p' \mid p\in L\}$.
\end{prop}

%[[ we can also write it in second-order propositional logic. ]]

The definition of $\Delta$ is similar to the definition of $F^*$ used in the second-order logic based definition of a stable model from~\cite{ferraris11stable}. This leads to the following reformulation of $\lpmln$ in second-order logic. 

Let ${\bf p}$ be a list of distinct atoms,
$p_1,\dots,p_n$, and let ${\bf u}$ be a list of distinct propositional variables $u_1,\dots,u_n$. 
By ${\bf u}\leq{\bf p}$ we denote the conjunction of the formulas
$\forall {\bf x}(u_i({\bf x})\rar p_i({\bf x}))$ for all $i=1,\dots n$,
where ${\bf x}$ is a list of distinct object variables whose length is
the same as the arity of $p_i$.
Expression ${\bf u}<{\bf p}$ stands for 
\hbox{$({\bf u}\leq{\bf p})\land\neg({\bf p}\leq{\bf u})$}.

\begin{prop} \label{prop:sorder}
For any $\lpmln$ program $\mo{F}$, a set $X$ of atoms is a soft stable model of $\mo{F}$ iff $X$ satisfies 
\[
  \neg\exists {\bf u} ({\bf u}<{\bf p})\land \Delta_{{\bf u}}(\{\o{\mo{F}}\}^{\rm ch}). 
\]

\end{prop}

%-----------------------------------------------------------------------------------------
\section{Checking Strong Equivalence Using ASP Solver} \label{sec:se-solver}
%-----------------------------------------------------------------------------------------

{Based on the {\bf Theorem on Soft Stable Models}, we use the following variant of Theorem~\ref{thm:se-char} to leverage an ASP solver for checking $\lpmln$ strong equivalence.
}

\medskip\noindent
{\bf Theorem~\ref{thm:se-char}$'$}\ \ 
{\sl 
For any $\lpmln$ programs $\mo{F}$ and $\mo{G}$, program $\mo{F}$ is strongly equivalent to $\mo{G}$ if and only if there is a w-expression $c_1+c_2\alpha$ such that for every interpretation $X$, 
\begin{enumerate}
%\item $TW(F\setminus \mo{F}_X) = c \times TW(G\setminus \mo{G}_X)$, and 
%\item (or equivalently, $P(\mo{F}_X) = c \times TW(\mo{G}_X)$, and 
\item[1a.]  
     $\sum\limits_{w:R\;\in\; \mo{F}, w\ne\alpha,\atop \text{ and } X\not\models R} w 
     \ \ =\ \  c_1 + \sum\limits_{w:R\;\in\; \mo{G}, w\ne\alpha, \atop \text{ and } X\not\models R} w$;
     
\item[1b.]     
     $|\{\alpha:R\;\in\; \mo{F}\ \mid\  X\not\models R \}|   =  c_2 +
     |\{\alpha:R\;\in\; \mo{G}\  \mid\  X\not\models R\} |$ ;
\item[2.] $\{p'\rar p \mid p \in {\bf p}\}$ entails
   $ \Delta_{\bf p'}(\{\o{\mo{F}}\}^{\rm ch})\lrar\Delta_{\bf p'}(\{\o{\mo{G}}\}^{\rm ch})$
(in the sense of classical logic).
\end{enumerate}
}\medskip

In each of the following subsections, we show how to check the conditions using {\sc clingo} together with {\sc f2lp} \cite{leej09}.  We need  {\sc f2lp} to turn propositional formulas under the stable model semantics into the input language of {\sc clingo}. 
We assume weights are given in integers as required by the input language of {\sc clingo}.

%-----------------------------------------------------------------------------------------
\subsection{Checking Conditions 1a, 1b of Theorem~\ref{thm:se-char}$'$}

In order to check Conditions~1(a),1(b) of Theorem~\ref{thm:se-char}$'$, we start by finding potential values for $c_1$ and $c_2$. For that, we arbitrarily set $X=\emptyset$ and find the values. 
%Notice that $\emptyset$ is always a soft stable model of an $\lpmln$ program, so its weight is non-zero. 
If the same values of $c_1$ and $c_2$ make the equations true for all other interpretations as well, the conditions hold.

The checking is done by using the program ${\bf P}$ in the input language of {\sc f2lp}, constructed as follows. 
For any soft rule $w_i: R_i$ in $\mo{F}$, where $w_i$ is an integer, ${\bf P}$ contains
\beq
\ba {rcl}
   {\tt f\_unsat\_s}(w_i, i) &\ar& {\tt not}\ R_i \\
   R_i &\!\!\ar\!\!& {\tt not}\ {\tt f\_unsat\_s}(w_i,i)
\ea
\eeq{p-soft}
and for any hard rule
$\alpha: F$ in $\mo{F}$, ${\bf P}$ contains 
\beq
\ba {rcl}
 {\tt f\_unsat\_h}(i) &\ar & {\tt not}\ R_i \\
  R_i & \!\!\ar\!\!& {\tt not}\ {\tt f\_unsat\_h}(i) .
 \ea
\eeq{p-hard}
Or if $R_i$ is already in the form 
\begin{center}
$\i{Head}_{i} \leftarrow \i{Body}_{i}$
\end{center}
allowed in the input language of {\sc clingo}, instead of \eqref{p-soft}, we can also use \footnote{
 In the case $\i{Head}_i$ is a disjunction $l_1; \cdots; l_n$, expression $\no\ \i{Head}_i$ stands for $\no\ l_1, \cdots, \no\ l_n$.}
\[
\ba{rcl}
{\tt f\_unsat\_s}(w_{i}, i) &\ar & \i{Body}_{i}, {\tt not}\ \i{Head}_{i}  \\
\i{Head}_{i} &\ar &  {\tt not}\ {\tt f\_unsat\_s}(w_{i}, i), \i{Body}_{i}
\ea
\]
and instead of \eqref{p-hard}, 
%sat_f(w_{i}, i) \leftarrow \neg \ unsat_f(w_{i}, i)\\
%for each rule R with soft weight in F, \\
\[
\ba {rcl}
{\tt f\_unsat\_h}(1, i) &\ar & \i{Body}_{i}, \no\ \i{Head}_{i}  \\
\i{Head}_{i} &\ar & {\tt not}\ {\tt f\_unsat\_h}(1, i), \i{Body}_{i}.
%sat_f(w_{i}, i) \leftarrow \neg \ unsat_f(w_{i}, i)\\
\ea
\]
${\bf P}$ contains similar rules for each weighted formula in $\mo{G}$ using
${\tt g\_unsat\_s}(\cdots)$ and  ${\tt g\_unsat\_h}(\cdots)$ atoms, as well as 
\[
{%\footnotesize
\ba l
{\tt f\_pw\_s}(S) \leftarrow 
    S=\#sum\{X,Y:  {\tt f\_unsat\_s}(X,Y),Y=1..i_f\} \\
{\tt g\_pw\_s}(S) \leftarrow 
    S=\#sum\{X, Y: {\tt g\_unsat\_s}(X,Y), Y=1..i_g\} \\
{\tt f\_pw\_h}(S) \leftarrow 
    S=\#count\{W:  {\tt f\_unsat\_h}(W),W=1..i_f\} \\
{\tt g\_pw\_h}(S) \leftarrow 
    S=\#count\{W: {\tt g\_unsat\_h}(W),W=1..i_g\}.
%}
\ea
}
\]
($i_f$ is the total number of rules in $\mo{F}$, and $i_g$ is the total number of rules in $\mo{G}$), and furthermore,
\beq
\neg p
\eeq{empty}
for each atom $p$ in ${\bf p}$ to ensure that we consider $X=\emptyset$.
%[[ there is a unique stable model here ]]

For example, for $\mo{F}$ and $\mo{G}$ in Example \ref{ex:main}, ${\bf P}$ is 
%which gives one stable model indicating the penalty weight of soft rule $\mo{F}$ and $\mo{G}$ by the value of the variable $x_1$ and $x_2$ in ${\tt f\_pw\_s}(x_1)$ ${\tt g\_pw\_s}(x_2)$ respectfully the as expected.  

%{\cred [[ @Man please update the program ]]}
\begin{lstlisting}
not a:- not f_unsat_s(0,1).
f_unsat_s(0,1):- not not a.
a :- not not a, not f_unsat_s(3, 2).
f_unsat_s(3,2):-not not a, not a. 
b:- a, not f_unsat_s(2, 3).
f_unsat_s(2, 3):- a, not b.
not a | b :- not g_unsat_s(2, 1).
g_unsat_s(2, 1):- not not a, not b.
a :- not not a, not g_unsat_s(1,2).
g_unsat_s(1,2) :- not not a, not a.
f_pw_s(S) :- S = #sum{X, Y: f_unsat_s(X, Y), Y=1..3}.
g_pw_s(S) :- S = #sum{X, Y: g_unsat_s(X, Y), Y=1..2}.
not a.
not b.
\end{lstlisting}

${\bf P}$ has a unique answer set, which tells us the potential parameters $c_1$ and $c_2$ for Conditions 1a and 1b each. If the answer set contains 
$\{{\tt f\_pw\_s}(x_1), 
{\tt f\_pw\_h}(x_2),
{\tt g\_pw\_s}(y_1),
{\tt g\_pw\_h}(y_2)
\}$
then let 
%Based on the output of this program, the parameters $c_1, c_2$ in $c$ can be computed by,
\begin{eqnarray*}
    c_1 = x_1 - y_1 \ \ \ \ \ \ \ 
    c_2 = x_2 - y_2. 
\end{eqnarray*}

Below we show how to check Condition~1 given $c_1$ and $c_2$ computed as above.
Let $\bf{P}^*$ is the program obtained from ${\bf P}$ by removing rules \eqref{empty}  for all atom $p\in{\bf p}$ and adding the following rules
\[
\ba l
    \leftarrow {\tt f\_pw\_s}(X), {\tt g\_pw\_s}(Y), X= Y + c_1  \\
    \leftarrow {\tt f\_pw\_h}(X), {\tt g\_pw\_h}(Y)), X= Y + c_2.  
\ea
\]

{\begin{prop} \label{prop:condition1}
Conditions~1a, 1b of Theorem~\ref{thm:se-char}$'$ hold iff 
$\bf{P}^*$ has no stable models.
\end{prop}
}
%-----------------------------------------------------------------------------------------
\subsection{Checking the second condition of Theorem~\ref{thm:se-char}$'$}
%-----------------------------------------------------------------------------------------
We check the second condition of Theorem~\ref{thm:se-char}$'$ by checking if each of the following ASP program is unsatisfiable.
Let ${\bf p}$ be the set of all atoms occurring in $\mo{F}$ and $\mo{G}$.

\noindent
${\bf P}^{**}_{1}$ is the following set of rules:
\begin{eqnarray*}
 &\{\{p\}^{\rm ch} \mid p \in {\bf p} \} \cup 
  \{\{p'\}^{\rm ch} \mid p \in {\bf p} \} \cup
 \{p' \rightarrow p \mid p \in {\bf p}\} 
 \cup\ \Delta_{{\bf p'}}(\{\o{\mo{F}}\}^{\rm ch}) \cup \neg\Delta_{{\bf p'}}(\{\o G\}^{\rm ch}).\nonumber
\end{eqnarray*}
${\bf P}^{**}_{2}$ is the following set of rules:
\begin{eqnarray*}
  &\{ \{p\}^{\rm ch} \mid p \in {\bf p} \} \cup
       \{ \{p'\}^{\rm ch} \mid p \in {\bf p} \}  \cup 
       \{p' \rightarrow p \mid p \in {\bf p}\}  
  \cup\ \Delta_{{\bf p'}}(\{\o G\}^{\rm ch}) \cup\
       \neg\Delta_{{\bf p'}}(\{\o{\mo{F}}\}^{\rm ch}).
\end{eqnarray*}

\medskip

%We use the equivalence between Condition (b) and Condition (j) of {\bf Theorem on Soft Stable Models} to check if $\mo{F}\cup \mo{H}$ and $\mo{G}\cup \mo{H}$ have the same set of stable models. 

%[[ {\cred @man: please update the program ]]}
%\begin{example}
For example, for $\mo{F}$ in Example~\ref{ex:main}, ${\bf P}^{**}_{1}$ in the input language of {\sc f2lp} is as follows.
%and $P^{**}_{2}$ in the input language of {\sc f2lp} is as follows. 
%$P^{**}_{1}$ is 
\begin{lstlisting}
{a; aa; b; bb}. 
aa -> a.
bb -> b.

% \Delta({F}^{ch})
not a | not not a.
(aa -> bb) & ( a-> b)| not ( a-> b).
(not not a->aa) & (not not a->a) | not (not not a->a).

% not \Delta({G}^{ch})
not ((not a | bb | not (not a | b)) & ((aa| not a) | not (a | not a))).
\end{lstlisting}

%\end{example}

\begin{prop} \label{prop:condition2}
% Condition 2 of Theorem~\ref{thm:se-char}$'$ is true iff both $\mo{P_1}^{**}$ and $\mo{P_2}^{**}$ have no stable models.
{Condition 2 of Theorem~\ref{thm:se-char}$'$ is true iff neither ${\bf P}_1^{**}$ nor ${\bf P}_2^{**}$ has stable models.}
\end{prop}

The structural equivalence checking method is related to the strong equivalence checking method using SAT solvers in \cite{chen05selp}. Paper \cite{janhunen04lpeq} reports another system for automated equivalence checking.

%-----------------------------------------------------------------------------------------
\section{Conclusion}
%-----------------------------------------------------------------------------------------
In this paper, we defined the concept of strong equivalence for $\lpmln$ programs and provide several equivalent characterizations. On the way, we have presented a few reformulations of $\lpmln$ that give us useful insight. 

The strong equivalence checking in Section~\ref{sec:se-solver} restricts soft rules' weights to integers only. We expect that this restriction can be removed if we use an external function call in {\sc clingo}.

Building upon the results presented here, we plan to extend the work to approximate strong equivalence, where the probability distributions may not necessarily be identical but allowed to be slightly different with some error bounds. This would be more practically useful for $\lpmln$ programs whose weights are learned from the data \cite{lee18weight}. 

\medskip\noindent
{\bf Acknowledgements:} 
We are grateful to the anonymous referees for their useful comments. This work was partially supported by the National Science Foundation under Grant IIS-1815337.

\bibliographystyle{eptcs}
%\bibliography{bib,bib2}

%\include{lpmln-equiv-iclp-cr-appendix-0728}

\end{document}